\pdfoutput=1
\documentclass[aps,prl,reprint,groupedaddress,floatfix]{revtex4-2}

\usepackage{graphicx}
\usepackage{dcolumn}
\usepackage{bm}
\usepackage{physics}
\usepackage{amssymb}
\usepackage{natbib}

\begin{document}


\title{Simple measurement of silanol (OH) concentration in silanol-terminated silicone fluids}


\author{Louis A. Bloomfield}
\email{lab3e@virginia.edu}
\affiliation{Department of Physics, University of Virginia, Charlottesville, VA 22904}


\date{\today}

\begin{abstract}
This article reports a simple technique for measuring the concentration of silanol groups in a silanol-terminated polydimethylsiloxane fluid. That technique requires only a common drying agent and a Zeolite, can be completed in minutes, and is accurate to about 10\%.

\end{abstract}


\maketitle

\section{Introduction}

Silanol-terminated polydimethylsiloxane fluids (STPDMS) are common commercial products that are most often sold by their viscosities. Those viscosities usually range from about 30 cSt to 50,000 cSt and correspond to ranges in molecular weights and silanol (hydroxyl) contents. In an ideal monodispersed STPDMS, the fluid's viscosity, molecular weight, and silanol content are directly related. However, commercial fluids are not monodispersed chemicals; instead they are products that generally contain statistical distributions of molecular weights. Moreover, some commercial fluids are blends of higher-viscosity and lower-viscosity STPDMS fluids, mixed together to achieve a particular target viscosity rather than a particular silanol concentration. As a result of inevitable variations in production and the possibility of blending in some products, different lots of the same product from the same vendor may differ significantly in silanol concentration.

It would therefore be useful to be able to measure silanol concentration in a simple way, one that does not require expensive equipment or the services of a commercial testing facility. Toward that end, we report such a simple technique. This technique uses a small amount of the STPDMS fluid, along with a common drying agent (trimethyl borate or TMB), and fresh 4A Zeolite powder.

\begin{figure}
	\includegraphics{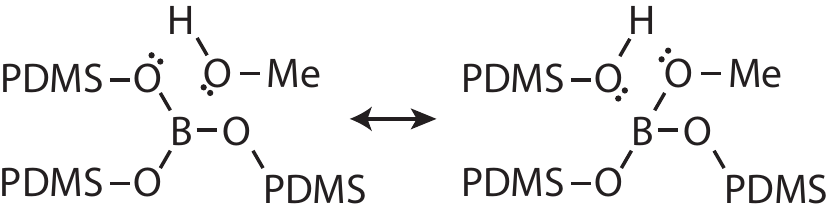}%
	\caption{The transesterification mechanism responsible crosslinking dynamics in mixtures of TMB and STPDMS. When an -OH group on either MeOH or STPDMS aligns with a =BO- crosslink, the lone electron pairs on the two oxygen atoms facilitate a rearrangement of bonds. The previously unattached molecule forms a covalent bond with the boron and a previously attached ligand is released with a new -OH group.\label{fig:exchangeReaction}}
\end{figure}

As observed by Bloomfield\cite{bloomfield2018}, TMB forms crosslinks between STPDMS chains by transesterification with the silanol groups (Fig. \ref{fig:exchangeReaction}). In each transesterification reaction, a silanol group attacks the boron atom of the TMB molecule, ejecting a methanol (MeOH) molecule and leaving the PDMS chain covalently bound to the boron atom. That reaction is reversible, meaning that a MeOH molecule can also attack the boron atom, ejecting an STPDMS chain and leaving the MeOH covalently bound to the boron atom as a methoxy group. These transesterification reactions are nearly barrier-less and proceed rapidly at room temperature.

In the continuing presence of STPDMS and MeOH molecules, the transesterification reactions proceed in both directions until they reach equilibrium. The rapid interchange of STPDMS chains and MeOH molecules on the boron atoms leaves the mixture liquid-like, despite it having a substantial number of crosslinks at any given time. To drive the reaction in one direction, one can use Le Chatelier's principle and remove one of the reaction products. If the MeOH is removed, then transesterification proceeds in the direction that links STPDMS chains to boron atoms until either all of the TMB or all of the STPDMS is consumed. When 4A Zeolite is present in sufficient quantity to adsorb all of the MeOH molecules produced by transesterification, the reaction is driven in one direction until it achieves maximum boron crosslinking. If the resulting crosslinking concentration exceeds a certain threshold, the mixture will become solid.

Because each boron atom can bond to 3 STPDMS chains, TMB is a crosslinker with coordination number 3. An STPDMS fluid will therefore be fully crosslinked or ``saturated'' when it has one boron atom crosslink for every three silanol groups in the original STPDMS fluid. TMB has a formula weight of 103.91 g/mol and can replace 3 silanol groups. The formula weight of TMB's methoxy groups, each of which can replace one silanol group) is thus 34.64 g/mol. Since OH has a formula weight of 17.01 g/mol, saturation occurs at $34.64/17.01$ or 2.04 wt\% TMB per wt\% OH in the original STPDMS. For example, a 70 cSt STPDMS fluid that contains 1.25 wt\% OH can be saturated (fully crosslinked) by $2.04\cdot1.25$ wt\% or 2.55 wt\% TMB.

However, the crosslinking reaction between TMB and STPDMS will cause the fluid to gel long before it is fully crosslinked. Gelation occurs when crosslinks are sufficiently concentrated to cause the fluid's polymer networks to transition from microscopic (molecular scale) to macroscopic (container scale)\cite{flory1941a,flory1941b,stockmayer1944}. For a coordination number 3 crosslinker, the gelation threshold occurs at 50\% of crosslink saturation. Thus if TMB is added slowly to an STPDMS fluid and the transesterification reaction is driven in the crosslinking direction, the fluid will gel once TMB weighing about $2.04\cdot0.5$ or 1.02 times the weight of OH in the original STPDMS fluid has been added. 

\section{Measurement Technique}

Our technique is based on that gelation effect. Specifically, we measure the amount of TMB required to gel a known amount of STPDMS in the presence of 4A Zeolite. To perform this measurement, a magnetic stirrer and 4A Zeolite powder are placed in a disposable plastic beaker. The amount of 4A Zeolite is at least enough to adsorb all the MeOH released during the crosslinking reaction that follows. A measured weight of STPDMS fluid is added to the beaker and the beaker is magnetically stirred until any residual moisture in the STPDMS fluid has been adsorbed into the Zeolite.

Prior to the addtion of TMB, the beaker is carefully pre-weighed. TMB is then added dropwise to the beaker's contents, stirring long enough after each drop for all of the MeOH released by the transesterification reaction to be adsorbed into the Zeolite---typically 15 seconds per drop. As the boron crosslink concentration approaches half of saturation, the liquid in the beaker will approach the gelation threshold and the liquid will become noticeably thicker. The rate of TMB addition must be slowed to ensure complete mixing and thorough adsorption of the MeOH by the Zeolite. When the gelation threshold is crossed, the liquid in the beaker will gel and the magnetic stirrer will stop spinning.

The beaker is then post-weighed and the amount of TMB required to cause gelation is calculated. The ratio of that TMB weight to the STPDMS fluid weight is then divided by 1.02 to obtain the approximate silanol weight fraction in the original STPDMS fluid.

Here we observe that to estimate the amount of 4A Zeolite required in the measurement, note that 4A Zeolite can adsorb up to about 19\% of its weight in MeOH\cite{walker1966}. TMB releases 93\% of its weight in MeOH during the measurement, so the weight of TMB need to reach geleation is approximately the weight of silanol group in the STPDMS fluid. The amount of 4A Zeolite needed to drive the crosslinking reaction to completion is thus at least 0.93/0.19 or about 5 times the weight of silanol groups in the STPDMS fluid. 

\section{Measurement Example}

This example measured the OH concentration of Dystar SFR 70 Lot MY26879, an STPDMS with a reported viscosity of 74.1 cPs at@ 25°C. Since the OH concentration of a typical 70 cSt STPDMS fluid is about 1.25\%, a 20g portion of that SFR 70 can be estimated to contain about $0.0125\cdot20$g or 0.250g of OH and require approximately $1.02\cdot0.250$g or 0.255g of TMB to reach the gelation threshold.

In preparation for the actual measurement, it was estimated that at least $5\cdot0.255$g or 1.28g of 4A Zeolite powder are needed to absorb the MeOH released by the transesterification reaction. As a margin of safety, 2.50g of 4A Zeolite powder were added the a 50 mL plastic beaker containing a magnetic stirrer.

20.003g of Dystar SFR 70 were added to the beaker, the 4A Zeolite was dispersed into that fluid, and the beaker was carefull pre-weighed. While the stirrer was spinning rapidly, producing a deep vortex in the cloudy fluid, TMB was added dropwise to the center of the vortex. About 15 seconds were allowed between drops, with each drop containing approximately 0.008g of TMB.

Once the amount of added TMB exceeded 0.200g, the time allowed between drops was increased to 30 seconds or longer. Finally, the fluid thickened to the point where the beaker begin to spin with the stirrer. It was not yet obvious that the fluid had gelled, but letting the beaker to sit for 5 minutes proved that the gelation threshold had been crossed. Post-weighing the beaker showed that 0.244g of TMB had been added.

The weight of OH in the original 20g of SFR 70 was therefore about $0.244$g$/1.02$ or 0.239g. That corresponds to about 1.20\% OH in the STPDMS fluid. Although the exact amount of TMB required reach the gelation threshold was probably a fraction of a drop less than 0.244g, this measurement is almost certainly accurate to within 10\%. 

\bibliography{silanol-measurement}

\end{document}